\newif\iftwocol
\twocolfalse
\iftwocol
  \documentclass[final,5p,times,twocolumn]{elsarticle}
\else
  \documentclass[final,3p,times]{elsarticle}
\fi

\usepackage{amsmath,amssymb,amsthm,mathtools,bm,booktabs,graphicx,microtype}
\usepackage{multirow}
\usepackage[hidelinks]{hyperref}
\hypersetup{pdftitle={Reference-Filter-Driven Transition Probabilities for IMM-Based Satellite Maneuver Detection},pdfauthor={Euiseok Han, Sangheon Choi, Seung-Hyun Kong}}
\usepackage{siunitx}
\iftwocol
  \usepackage{placeins}
  
\else
  \usepackage[section]{placeins}
\fi
\makeatletter
\def\ps@pprintTitle{\let\@oddhead\@empty\let\@evenhead\@empty
  \def\@oddfoot{\hfil\thepage\hfil}\let\@evenfoot\@oddfoot}
\makeatother

\newcommand{\earlyfloat}[1]{\iftwocol#1\fi}
\newcommand{\latefloat}[1]{\iftwocol\else#1\fi}
\newlength{\colfigwidth}
\iftwocol
  \newenvironment{widefigure}[1][!t]{\begin{figure*}[#1]}{\end{figure*}}
  \newenvironment{widetable}[1][t]{\begin{table*}[#1]}{\end{table*}}
\else
  \newenvironment{widefigure}[1][!tb]{\begin{figure}[#1]}{\end{figure}}
  \newenvironment{widetable}[1][!tb]{\begin{table}[#1]}{\end{table}}
\fi
\graphicspath{{figures/}}

\usepackage{xcolor}
\newcommand{\rv}{\bm{r}}
\newcommand{\vv}{\bm{v}}
\newcommand{\av}{\bm{a}}
\newcommand{\xv}{\bm{x}}
\newcommand{\zv}{\bm{z}}
\newcommand{\hv}{\bm{h}}
\newcommand{\nuv}{\bm{\nu}}
\newcommand{\dv}{\Delta\bm{v}}
\newcommand{\dvs}{\Delta v}

\newcommand{\Pca}{\mu^{\mathrm{M}}}

\newcommand{\Psiref}{\Psi^{\mathrm{ref}}}
\newcommand{\pon}{p_{\mathrm{CM}}}
\newcommand{\ponk}{p_{\mathrm{CM},k}}
\newcommand{\poff}{p_{\mathrm{MC}}}
\newcommand{\poffk}{p_{\mathrm{MC},k}}
\newcommand{\rdimm}{RD-IMM}

\newcommand{\nMC}{50}
\newcommand{\nCstar}{3.87}
\newcommand{\nPropPdTone}{0.80}
\newcommand{\nPropPdZone}{0.98}
\newcommand{\nPropLatTone}{107}
\newcommand{\nPropLatZone}{42}
\newcommand{\nPropFaNull}{0.03}
\newcommand{\nPropFaNullFifty}{0.02}
\newcommand{\nJiaFa}{0.11}
\newcommand{\nGateFa}{0.04}
\newcommand{\nOwnFa}{0.02}

\newcommand{\nGatePdTone}{0.56}
\newcommand{\nGateLatTone}{128}
\newcommand{\nGateLatZone}{45}

\newcommand{\nOwnPdTone}{0.02}
\newcommand{\nJiaPdTone}{0.06}
\newcommand{\nGoffPdTone}{0.56}

\newcommand{\nPropRmseCoast}{0.74}

\newcommand{\nJiaRmseCoast}{1.09}

\newcommand{\nNtestGeo}{371}

\newcommand{\nPropPdThalf}{0.12}
\newcommand{\nPropPdZhalf}{0.36}

\newcommand{\nXieFa}{0.02}
\newcommand{\nXiePdTone}{0.00}
\newcommand{\nXiePdZone}{0.02}

\journal{Acta Astronautica}

\begin{document}

\begin{frontmatter}

\title{Reference-Filter-Driven Transition Probabilities for IMM-Based
Satellite Maneuver Detection}

\author[a]{Euiseok Han}
\ead{hanes1207@kaist.ac.kr}
\author[b]{Sangheon Choi}
\ead{sangheon_choi@kaist.ac.kr}
\author[c]{Seung-Hyun Kong\corref{cor1}}
\ead{skong@kaist.ac.kr}
\cortext[cor1]{Corresponding author}
\address[a]{School of Electrical Engineering, Korea Advanced Institute of Science and Technology (KAIST), 291 Daehak-ro, Yuseong-gu, Daejeon 34141, Republic of Korea}
\address[b]{Department of Aerospace Engineering, Korea Advanced Institute of Science and Technology (KAIST), 291 Daehak-ro, Yuseong-gu, Daejeon 34141, Republic of Korea}
\address[c]{Cho Chun Shik Graduate School of Mobility, Korea Advanced Institute of Science and Technology (KAIST), 193 Munji-ro, Yuseong-gu, Daejeon 34051, Republic of Korea}

\begin{abstract}
Detecting maneuvers of non-cooperative satellites is essential for space situational awareness.
The interacting multiple model (IMM) filter has been applied to detect unannounced maneuvers, but the standard IMM assumes a fixed Markov transition probability matrix (TPM) that must be tuned without knowledge of the maneuver rate of the target.
To avoid the tuning, adaptive-TPM approaches update the TPM online from statistics computed inside the IMM.
However, the statistics already depend on the TPM through the interaction step, and the dependence forms a feedback loop.
The loop itself can raise the maneuver probability in the absence of a maneuver or suppress its rise when a maneuver occurs.
In this paper, the TPM is driven by the normalized innovation squared (NIS) of a reference coast filter that is decoupled from the IMM, and the NIS is mapped continuously to the coast-to-maneuver transition probability.
The reference-driven IMM (\rdimm) is evaluated with four space-based optical sensors tracking a geostationary target.
In the simulation, the reference filter keeps its sensitivity after small maneuvers, whereas the NIS of the coast sub-filter of the IMM is reduced by the interaction step.
\rdimm{} detects 1--3~m/s maneuvers, which the fixed-TPM IMM rarely detects, with a false-declaration rate below the false-alarm probability of 0.05 set for the reference filter.
In the first hour after the 1--3~m/s maneuvers, \rdimm{} also has a lower position error than the compared filters.
\end{abstract}


\end{frontmatter}

\section{Introduction}\label{sec:intro}

Detecting the maneuvers of non-cooperative satellites is an important task of space situational awareness~\cite{kelecy2007satellite,lemmens2014two,patera2008space,holzinger2012object,bobrinsky2010space}, because the risk of collision increases as space becomes congested~\cite{weeden2010global}.
Conjunction screening for collision avoidance relies on orbit determination solutions, which do not capture a maneuver unless the operator provides an ephemeris that includes the maneuver~\cite{nasa2023ca}.
Since satellites are operated by different organizations, many of them are non-cooperative and do not announce their maneuvers.
Geostationary satellites, for example, perform station-keeping maneuvers every one or two weeks~\cite{woo2021geostationary,pastor2022satellite}.
When an unannounced maneuver corrupts the orbit estimate, new observations of the satellite may fail to correlate with the catalogued orbit~\cite{hill2014maneuver,pastor2022satellite}, and a conjunction with the satellite may not be predicted from the corrupted estimate~\cite{nasa2023ca}.
Therefore, a maneuver should be detected shortly after it occurs while false declarations are kept rare, because a false declaration may also degrade the orbit estimate of a satellite that has not maneuvered.

Currently, operational catalogues are maintained with single-model estimators such as weighted batch least squares~\cite{nasa2023ca,velasco2023generic}, and maneuvers are detected from outliers in the measurement residuals~\cite{velasco2023generic}.
Batch and tracklet-based methods detect geostationary maneuvers of a few cm/s from ground optical data~\cite{hill2014maneuver,pastor2022satellite}.
However, consecutive tracks of an object are separated by hours to days, and a maneuver is detected only after the next track~\cite{hill2014maneuver}.
In contrast, a high-interest geostationary object can be observed intensively by a network of sensors~\cite{george2021phantom}, and multiple space-based optical sensors can track it continuously~\cite{jia2015space}.
When the object is tracked continuously, the tracking filter itself can detect the maneuver from the measurement stream.

An orbit determination filter with a single dynamics model diverges when the spacecraft maneuvers without notice~\cite{goff2015tracking}.
To address the divergence, the interacting multiple model (IMM) filter~\cite{blom1988interacting,li2005survey,bar2001estimation} has been used to track non-cooperative satellites through their maneuvers~\cite{jia2015space,goff2015tracking} and to detect the maneuvers~\cite{jia2015space}.
A two-model IMM describes the target dynamics with a coast model and a maneuver model and combines their estimates with mode probabilities, which are updated from the model likelihoods and a Markov transition probability matrix (TPM).
Conventionally, the TPM is fixed in advance~\cite{jia2015space,lee2023improved,xie2019adaptive}, which assumes a constant maneuver rate, whereas the maneuvers of a non-cooperative satellite are sparse and irregular.
A small coast-to-maneuver transition probability $\pon$ keeps the maneuver probability low during coast and preserves the coast accuracy, but delays the switch to the maneuver mode after a maneuver, or prevents the switch after a small maneuver.
A large $\pon$ makes the switch faster and enables the switch after a small maneuver, at the cost of a higher maneuver probability, more frequent false declarations and a larger estimation error during coast.
The choice of $\pon$ is therefore a trade-off between coast accuracy and responsiveness to a maneuver~\cite{mondal2026maneuver}, and the optimal value for the trade-off depends on the maneuver rate, which is unknown for a non-cooperative satellite.

To mitigate the trade-off, adaptive-TPM IMM filters update the TPM online from the mode probabilities or model likelihoods of an IMM~\cite{jilkov2004online,lee2023improved,xie2019adaptive,chen2024model,zhang2023orbit,mondal2026maneuver,zhang2026hybrid}.
Jilkov and Li~\cite{jilkov2004online} estimate the TPM recursively in a Bayesian framework from the model likelihoods and mode probabilities of the IMM.
Lee and Park~\cite{lee2023improved} correct the TPM with functions of the posterior mode probabilities, and Chen et al.~\cite{chen2024model} combine the correction with a model-switching detection.
Xie et al.~\cite{xie2019adaptive} run a fixed-TPM IMM in parallel with an adaptive IMM, and correct the TPM of the adaptive IMM with the mode-probability increments of the fixed-TPM IMM when a model-jump statistic falls below a threshold.
The mode probability of the fixed-TPM IMM does not depend on the adapted TPM, but the fixed-TPM IMM inherits the trade-off of the fixed TPM, and the adaptive IMM can therefore mitigate the trade-off only partly.
Zhang et al.~\cite{zhang2023orbit} adjust the TPM from the mode probabilities to track a spacecraft with continuous thrust from angle-only measurements, and Mondal et al.~\cite{mondal2026maneuver} update the TPM from the per-model likelihoods of the sub-filters to avoid the delay induced by the maneuver detection from a moving average of the normalized innovation squared (NIS).
Zhang et al.~\cite{zhang2026hybrid} generate the TPM with a neural network from the mode-probability increments, model likelihood ratios and acceleration estimates of the IMM for the relative navigation of a non-cooperative spacecraft.
However, the studies of adaptive-TPM IMM filters evaluate trajectories with maneuvers or mode switches and do not quantify the rate of false declarations in maneuver-free runs.

When the TPM is driven by statistics computed inside the IMM, a feedback loop is formed through the TPM, and the loop can lead to false declarations and missed detections.
In the interaction step, the TPM and the mode probabilities determine the predicted mode probabilities and the mixed initial state estimates of the mode-matched filters.
Every statistic computed inside the IMM therefore recursively depends on the TPM of the previous time steps.
Depending on the driving statistic~\cite{lee2023improved,xie2019adaptive,chen2024model,zhang2023orbit,jilkov2004online,mondal2026maneuver} and the gain of the TPM update, the loop can raise the maneuver probability above the coast level without a maneuver, keep the TPM close to a fixed TPM after a small maneuver, or suppress the increase of $\pon$ after a maneuver (Section~\ref{sec:loop}).
When the TPM is driven by the NIS of the coast sub-filter, the mixing in the interaction step reduces the NIS after a maneuver.

The main contribution of this paper is to identify the feedback loop of adaptive-TPM IMM filters, which can lead to false declarations and missed detections when maneuvers are sparse.
To eliminate the loop, the TPM is driven by a statistic computed outside the IMM.
Goff et al.~\cite{goff2015tracking} also detect a maneuver when the NIS of a single-model filter exceeds a threshold and then switch to an IMM of covariance-inflated models (CI-IMM), but the TPM of the CI-IMM is fixed.
Motivated by the threshold-based detection of the CI-IMM, the proposed reference-driven IMM (\rdimm) takes the NIS from a reference coast filter that processes the same measurements but is decoupled from the IMM.
The NIS of the reference filter is mapped continuously to $\pon$ through the $\chi^2$ surprisal, a fading maximum and a logistic function.
Because the reference filter is decoupled from the IMM, its NIS does not depend on the TPM and approximately follows the $\chi^2$ distribution during coast, and the false-alarm probability of the reference filter can be set as a design value.
In contrast to the parallel fixed-TPM IMM of Xie et al.~\cite{xie2019adaptive}, the reference filter contains no TPM and does not inherit the trade-off of a fixed TPM.
In addition, since $\pon$ increases continuously with the NIS, a moderate NIS raises the predicted maneuver probability over consecutive measurements, and the IMM can declare a maneuver for which the NIS of the reference filter remains below the threshold of~\cite{goff2015tracking}.

\rdimm{} is compared with the IMM-CKF of Jia et al.~\cite{jia2015space} with a fixed TPM, the ATPM-PIMM of Xie et al.~\cite{xie2019adaptive} and the CI-IMM of Goff et al.~\cite{goff2015tracking} in a space-based optical scenario adapted from~\cite{jia2015space}.
In the scenario, \rdimm{} detects 1--3~m/s maneuvers, which include north--south station-keeping maneuvers of geostationary satellites~\cite{woo2021geostationary}, whereas the detection probabilities of the IMM-CKF and ATPM-PIMM for the 1~m/s maneuvers remain near their false-declaration rates.
The false-declaration rate of \rdimm{} remains below the false-alarm probability of 0.05 set for the reference filter.
In the first hour after the 1--3~m/s maneuvers, \rdimm{} also has a lower position error than the compared filters.

The remainder of this paper is organized as follows.
Section~\ref{sec:problem} formulates the orbital dynamics with coast and maneuver models, the angle-only measurement model and the IMM filter.
Section~\ref{sec:method} discusses the feedback loop of internal-statistics TPM adaptation and derives the reference-driven TPM.
Section~\ref{sec:sim} compares \rdimm{} with the three baseline filters in the space-based optical scenario, and
Section~\ref{sec:conclusion} concludes the paper.

\section{Problem formulation}\label{sec:problem}

\subsection{Orbital dynamics and maneuver model}\label{sec:dyn}
The satellite motion is modeled with two state-space models on the same 6-dimensional state, a coast model and a maneuver model that differs from it only in a larger process noise, as in the IMM of~\cite{jia2015space}.
All vectors are expressed in the Earth-centered inertial (ECI) frame. 
The position and velocity of the satellite are denoted by $\rv=[x,\,y,\,z]^{\top}$ and $\vv=\dot\rv$.
The state of both models is
\begin{equation}
    \xv^{j}=[\rv^{\top},\,\vv^{\top}]^{\top}\in\mathbb R^{6},\qquad j\in\{\mathrm C,\mathrm M\},
\end{equation}
where the superscripts C and M denote the coast and maneuver models.
The true equation of motion is
\begin{equation}
    \begin{aligned}
        \ddot\rv&=\nabla U(\rv)+\av_{3\mathrm b}(\rv,t)+\av_{\mathrm{srp}}(\rv,t)+\bm F,\\
        U(\rv)&=\frac{GM}{r}\Big[1-\sum_{n=2}^{6}J_n\Big(\frac{R_E}{r}\Big)^{\!n}P_n\Big(\frac{z}{r}\Big)\Big],
    \end{aligned}
    \label{eq:truth}
\end{equation}
where $U$ is the zonal gravity potential, $r=\|\rv\|$, $GM$ the Earth's gravitational parameter, $R_E$ its equatorial radius, $J_n$ the zonal harmonic coefficients, $P_n$ the Legendre polynomial of degree $n$, $\av_{3\mathrm b}$ the lunisolar third-body acceleration, $\av_{\mathrm{srp}}$ the solar radiation pressure acceleration for a spherical spacecraft with reflectivity coefficient $C_R$ and area-to-mass ratio $A/m$, computed with a cylindrical Earth shadow model, and $\bm F$ the true thrust acceleration.
The filters model the potential up to the $J_2$ term~\cite{vallado2001fundamentals},
\begin{equation}
    U_{J_2}(\rv)=\frac{GM}{r}-\frac{GM\, J_2R_E^2}{2r^3}\Big(\frac{3z^2}{r^2}-1\Big),
    \label{eq:PhiJ2}
\end{equation}
omitting the higher zonals and the third-body and radiation-pressure terms of Eq.~\eqref{eq:truth}.
The omitted terms act as unmodeled accelerations within the filters and are accounted for by the process noise.

During coast, the thrust acceleration vanishes ($\bm F=\bm 0$). 
The equation of motion of the coast model is
\begin{equation}
    \dot\xv^{\mathrm C}=f^{\mathrm C}(\xv^{\mathrm C})+\bm w^{\mathrm C}
    =\begin{bmatrix}\vv\\ \nabla U_{J_2}(\rv)\end{bmatrix}+\bm w^{\mathrm C},
    \label{eq:coast}
\end{equation}
where $\bm w^{\mathrm C}$ is a zero-mean white process noise with power spectral density $\tilde Q_{\mathrm C}$.
The filters integrate Eq.~\eqref{eq:coast} with a step of $\delta t=\SI{10}{s}$, using two fourth-order Runge--Kutta substeps per step.
A maneuver is modeled as an impulsive velocity change $\dv$ applied at an unknown time $t_b$, $\bm F=\dv\,\delta(t-t_b)$.
The maneuver model propagates the same dynamics, Eq.~\eqref{eq:coast}, and absorbs the velocity change through a larger discrete process noise $Q_{\mathrm M}$ (Section~\ref{sec:setup}) replacing the process noise of the coast model at every integration step.
A maneuver model with a 9-dimensional constant-acceleration state is also examined in Section~\ref{sec:sim}.

\subsection{Measurement model}\label{sec:measmodel}
The observation scenario follows Jia et al.~\cite{jia2015space}, with the target placed on the geostationary orbit as described in Section~\ref{sec:setup}.
The target is tracked by $n_s$ space-based optical sensors indexed by $q=1,\dots,n_s$.
The sensors provide angle-only measurements, which are the main measurement type for geosynchronous objects~\cite{siminski2014short,escribano2022automatic}.
At time $t_k=k\,\Delta t$ with the measurement interval $\Delta t=\SI{60}{s}$ used in~\cite{jia2015space}, sensor $q$ located at ECI position $\rv^{s}_{q}(t_k)$ measures the azimuth and elevation of the line-of-sight vector $\bm d_{q}=\rv-\rv^s_{q}(t_k)$ in the ECI frame:
\begin{equation}
    \begin{gathered}
        \zv_{k,q}=\hv_{q}(\rv(t_k))+\bm\epsilon_{k,q},\qquad
        \bm\epsilon_{k,q}\sim\mathcal N\big(\bm 0,\,\kappa_{q}\sigma_\theta^2I_2\big),\\
        \hv_{q}(\rv)=\begin{bmatrix}\tan^{-1}\big(d_{q,y}/d_{q,x}\big)\\[2pt]
        \tan^{-1}\Big(d_{q,z}\big/\sqrt{d_{q,x}^2+d_{q,y}^2}\Big)\end{bmatrix},
    \end{gathered}
    \label{eq:azel}
\end{equation}
where the azimuth is computed with the four-quadrant inverse tangent.
Here $\sigma_\theta$ is the standard deviation of each angle, and the noise covariance $\sigma_\theta^2I_2$ is scaled by a factor $\kappa_{q}$ that increases with the bistatic solar angle at the target. 
As modeled in~\cite{jia2015space}, $\kappa_{q}=0.8$, 0.9, 1.0, 1.1 and 1.2 for bistatic solar angles below \SI{20}{\degree}, \SI{40}{\degree}, \SI{60}{\degree}, \SI{75}{\degree} and \SI{90}{\degree}, respectively.

Sensor $q$ is considered available at $t_k$ if its line of sight is not blocked by the Earth and the bistatic solar angle is below \SI{90}{\degree}.
Let $\mathcal S_k=\{q_1,\dots,q_{m_k}\}$ denote the set of $m_k$ sensors available at $t_k$. 
The measurements from the $m_k$ sensors are concatenated into a single vector, whose measurement function and noise covariance are given by
\begin{equation}
    \begin{gathered}
        \zv_k=\big[\zv_{k,q_1}^{\top},\dots,\zv_{k,q_{m_k}}^{\top}\big]^{\top}\in\mathbb R^{2m_k},\\
        \hv_k(\xv^j)=\big[\hv_{q_1}(\rv)^{\top},\dots,\hv_{q_{m_k}}(\rv)^{\top}\big]^{\top},\\
        R_k=\sigma_\theta^2\,\mathrm{diag}\big(\kappa_{q_1}I_2,\dots,\kappa_{q_{m_k}}I_2\big).
    \end{gathered}
    \label{eq:meas}
\end{equation}

\subsection{Interacting multiple model filter}\label{sec:imm}
The coast and maneuver models are incorporated into an IMM filter~\cite{blom1988interacting,bar2001estimation}.
The mode transitions follow a Markov chain with the transition probability matrix (TPM)
\begin{equation}
        \Pi_k=\begin{pmatrix}1-\ponk&\ponk\\ \poffk&1-\poffk\end{pmatrix},\qquad
        p_{ij,k}=\Pr\{\theta_k=j\mid \theta_{k-1}=i\},\quad i,j\in\{\mathrm C,\mathrm M\},
    \label{eq:tpmdef}
\end{equation}
where $\theta_k$ is the mode in effect over $(t_{k-1},t_k]$, and $\ponk$ and $\poffk$ are the coast-to-maneuver and maneuver-to-coast probabilities.
Each IMM cycle consists of interaction, mode-matched filtering, mode-probability update and combination.
The mode probabilities are initialized to $\mu^{\mathrm C}_0=0.9$ and $\mu^{\mathrm M}_0=0.1$.

\paragraph{Interaction}
Given the mode probabilities $\mu^i_{k-1}$ and the mode-conditioned estimates $(\hat\xv^i_{k-1|k-1},P^i_{k-1|k-1})$, the predicted mode probability $\bar c^{j}_{k}$, the mixing probability $\mu^{i|j}_{k-1}$, and the mixed initial conditions for filter $j$ are given by
\begin{equation}
    \begin{aligned}
        \bar c^{j}_{k}&=\sum_i p_{ij,k}\,\mu^i_{k-1}, \qquad
        \mu^{i|j}_{k-1}=\frac{p_{ij,k}\,\mu^i_{k-1}}{\bar c^{j}_{k}},\\
        \hat\xv^{0j}_{k-1|k-1}&=\sum_i\mu^{i|j}_{k-1}\,\hat\xv^{i}_{k-1|k-1},\\
        P^{0j}_{k-1|k-1}&=\sum_i\mu^{i|j}_{k-1}\Big[P^{i}_{k-1|k-1}+\big(\hat\xv^{i}_{k-1|k-1}-\hat\xv^{0j}_{k-1|k-1}\big)\big(\hat\xv^{i}_{k-1|k-1}-\hat\xv^{0j}_{k-1|k-1}\big)^{\!\top}\Big].
    \end{aligned}
    \label{eq:mix}
\end{equation}

\paragraph{Mode-matched filtering}
Each mode-matched filter is implemented with a cubature Kalman filter (CKF)~\cite{arasaratnam2009cubature}, following the approach in~\cite{jia2015space}. 
With the state dimension $n=6$ of both models, the $2n$ cubature points for the mixed initial condition are given by
\begin{equation}
        \mathcal X^{(l)}_{k-1|k-1}=\hat\xv^{0j}_{k-1|k-1}+L^{0j}_{k-1|k-1}\bm\xi^{(l)},\qquad
        \bm\xi^{(l)}=\sqrt{n}\,[\bm 1]^{(l)},\quad l=1,\dots,2n,
    \label{eq:cub}
\end{equation}
where $L^{0j}_{k-1|k-1}$ is the lower-triangular Cholesky factor of $P^{0j}_{k-1|k-1}$, satisfying $P^{0j}_{k-1|k-1}=L^{0j}_{k-1|k-1}\big(L^{0j}_{k-1|k-1}\big)^{\top}$, and $[\bm 1]^{(l)}$ is the $l$-th column of the matrix $[I_{n},\,-I_{n}]$~\cite{arasaratnam2009cubature}. 
Each cubature point is propagated over one integration step $\delta t$ by numerically integrating Eq.~\eqref{eq:coast}, and the predicted state and covariance at the end of the step, denoted by the subscript $+$, are computed as
\begin{equation}
    \begin{aligned}
        \hat\xv^{j}_{+}&=\frac{1}{2n}\sum_l\mathcal X^{(l)}_{+},\\
        P^{j}_{+}&=\frac{1}{2n}\sum_l\big(\mathcal X^{(l)}_{+}-\hat\xv^{j}_{+}\big)\big(\mathcal X^{(l)}_{+}-\hat\xv^{j}_{+}\big)^{\!\top}+Q_{j}.
    \end{aligned}
    \label{eq:ckfpred}
\end{equation}
The discrete process noise covariance $Q_{\mathrm C}$ of the coast model over $\delta t$ is obtained from the power spectral density $\tilde Q_{\mathrm C}$ and the Jacobian $A_{\mathrm C}=\partial f^{\mathrm C}/\partial\xv$ evaluated at the predicted state by the Van Loan method~\cite{van1978computing,jia2015space}:
\begin{equation}
        \exp\!\left(\begin{bmatrix}-A_{\mathrm C}&\tilde Q_{\mathrm C}\\ 0&A_{\mathrm C}^{\top}\end{bmatrix}\delta t\right)
        =\begin{bmatrix}\Phi_{\mathrm C}^{-1}&\Phi_{\mathrm C}^{-1}Q_{\mathrm C}\\ 0&\Phi_{\mathrm C}^{\top}\end{bmatrix},\qquad
        Q_{\mathrm C}=\Phi_{\mathrm C}\big(\Phi_{\mathrm C}^{-1}Q_{\mathrm C}\big),
    \label{eq:vanloan}
\end{equation}
where $\Phi_{\mathrm C}$ is the state transition matrix over $\delta t$. 
For the maneuver model, the discrete covariance $Q_{\mathrm M}$ is specified directly, as in~\cite{jia2015space}.
The prediction of Eqs.~\eqref{eq:cub}--\eqref{eq:vanloan} is repeated $\Delta t/\delta t$ times between measurement times, and the cubature points of each step are generated from the predicted state and covariance of the previous step.
The last step gives $\hat\xv^j_{k|k-1}$ and $P^j_{k|k-1}$, from which the predicted points $\mathcal X^{(l)}_{k|k-1}$ are generated by Eq.~\eqref{eq:cub}.

For the measurement update, the predicted cubature points are propagated through the observation model in Eq.~\eqref{eq:meas}: 
\begin{equation}
    \begin{aligned}
        \mathcal Z^{(l)}&=\hv_k\big(\mathcal X^{(l)}_{k|k-1}\big),\qquad
        \hat\zv^j_{k}=\frac{1}{2n}\sum_l\mathcal Z^{(l)},\\
        \nuv^j_k&=\zv_k-\hat\zv^j_k,\\
        S^j_k&=\frac{1}{2n}\sum_l\big(\mathcal Z^{(l)}-\hat\zv^j_k\big)\big(\mathcal Z^{(l)}-\hat\zv^j_k\big)^{\!\top}+R_k,\\
        P^{j}_{xz}&=\frac{1}{2n}\sum_l\big(\mathcal X^{(l)}_{k|k-1}-\hat\xv^j_{k|k-1}\big)\big(\mathcal Z^{(l)}-\hat\zv^j_k\big)^{\!\top},\\
        K^j_k&=P^{j}_{xz}\big(S^j_k\big)^{-1},\qquad
        \hat\xv^j_{k|k}=\hat\xv^j_{k|k-1}+K^j_k\nuv^j_k,\\
        P^j_{k|k}&=P^j_{k|k-1}-K^j_kS^j_kK^{j\top}_k,
    \end{aligned}
    \label{eq:ckfupd}
\end{equation}
where the azimuth component of the innovation is normalized to $(-\pi,\pi]$. 
The innovation $\nuv^j_k$ and the associated covariance $S^j_k$ define the model likelihood and the normalized innovation squared (NIS) for filter $j$ as follows:
\begin{equation}
    \Lambda^j_k=\mathcal N\big(\nuv^j_k;\bm0,S^j_k\big),\qquad
    \Psi^j_k=\nuv^{j\top}_k\big(S^j_k\big)^{-1}\nuv^j_k .
    \label{eq:nis}
\end{equation}

\paragraph{Mode-probability update and combination}
The mode probabilities are updated with the likelihoods, and the combined state estimate is computed as follows:
\begin{equation}
    \mu^j_k=\frac{\Lambda^j_k\,\bar c^{j}_{k}}{\sum_i\Lambda^i_k\,\bar c^{i}_{k}}, \qquad
    \hat\xv_{k|k}=\mu^{\mathrm C}_k\hat\xv^{\mathrm C}_{k|k}+\mu^{\mathrm M}_k\hat\xv^{\mathrm M}_{k|k} .
    \label{eq:imm}
\end{equation}
A maneuver is declared at $t_k$ when the maneuver mode dominates,
\begin{equation}
    \mu^{\mathrm M}_k>0.5 .
    \label{eq:declare}
\end{equation}

\section{Reference-driven IMM}\label{sec:method}
In this section, the reference-driven IMM is presented. 
In the IMM, the initial state estimates of the coast and maneuver sub-filters, which are the mode-matched filters of Section~\ref{sec:imm}, are obtained by mixing their state estimates of the previous time step using mixing probabilities computed from the TPM and the mode probabilities.
Section~\ref{sec:loop} describes the feedback loop formed through the interaction step when the TPM is adapted from statistics computed inside the IMM, and the effects of the loop on the maneuver declaration.
Sections~\ref{sec:decoupled} and~\ref{sec:map} describe the proposed approach, in which the TPM is driven by a reference filter that uses the coast model of the IMM and is decoupled from the IMM.

\subsection{TPM adaptation driven by internal statistics}\label{sec:loop}
\begin{widefigure}[!t]\centering
    \includegraphics[width=\textwidth]{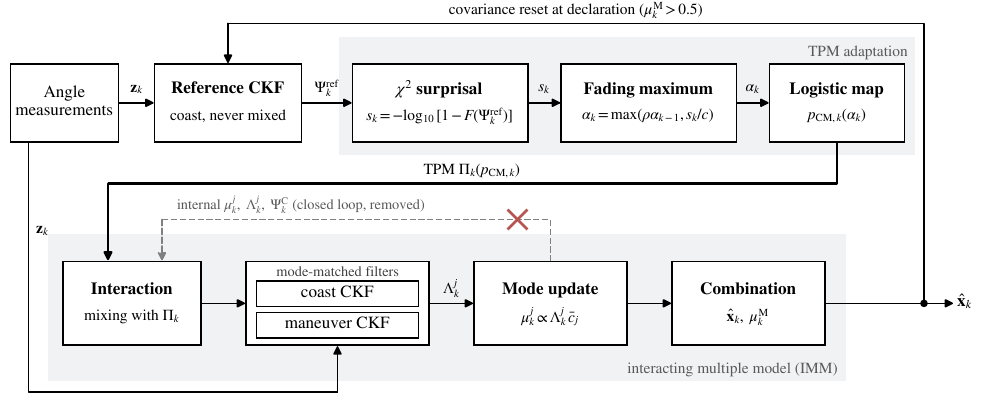}
    \caption{Structure of the reference-driven IMM (\rdimm).}
    \label{fig:arch}
\end{widefigure}
In adaptive-TPM IMM filters, the TPM is updated at each measurement time according to
\begin{equation}
    \Pi_k=\mathcal T\big(\Pi_{k-1},\,s_{k-1}\big),
    \label{eq:adapt}
\end{equation}
where $\mathcal T$ denotes an update mapping and $s_k$ is a driving statistic. 
In the literature, $s_k$ is typically selected as the posterior mode probability $\mu^j_k$, its increment $\mu^j_k-\mu^j_{k-1}$~\cite{xie2019adaptive,chen2024model,zhang2023orbit}, the current and previous mode probabilities~\cite{lee2023improved}, the model likelihood computed before mixing~\cite{mondal2026maneuver}, or the model likelihoods combined with the mode probabilities~\cite{jilkov2004online}.
In the adaptive-TPM methods, $s_k$ is computed inside the IMM whose TPM is adapted, except partly in the method of Xie et al.~\cite{xie2019adaptive}, which switches to the mode-probability increments of a parallel fixed-TPM IMM when a model-jump statistic falls below a threshold.
The TPM recursively affects the statistics computed inside the IMM in two ways through the interaction step, Eq.~\eqref{eq:mix}: it changes the predicted mode probabilities and the mixed initial state estimates.
First, the predicted probability of the maneuver mode,
\begin{equation}
    \bar c^{\mathrm M}_{k}=\ponk\,\mu^{\mathrm C}_{k-1}+\big(1-\poffk\big)\,\mu^{\mathrm M}_{k-1},
    \label{eq:cbarM}
\end{equation}
increases with $\pon$, and $\mu^{\mathrm M}_k$ increases with $\bar c^{\mathrm M}_{k}$ by Eq.~\eqref{eq:imm}.
Second, the mixed initial estimate of the coast sub-filter is
\begin{equation}
    \hat\xv^{0\mathrm C}_{k-1|k-1}=\hat\xv^{\mathrm C}_{k-1|k-1}+\mu^{\mathrm M|\mathrm C}_{k-1}\big(\hat\xv^{\mathrm M}_{k-1|k-1}-\hat\xv^{\mathrm C}_{k-1|k-1}\big),\qquad
    \mu^{\mathrm M|\mathrm C}_{k-1}=\frac{\poffk\,\mu^{\mathrm M}_{k-1}}{\bar c^{\mathrm C}_{k}},
    \label{eq:mixC}
\end{equation}
and the coast sub-filter is pulled toward the estimate of the maneuver sub-filter as $\mu^{\mathrm M}_{k-1}$ increases.
As a result, the innovation and likelihood $\Lambda^{\mathrm C}_k$ of the coast sub-filter also depend on $\mu^{\mathrm M}_{k-1}$.

Since $\mu^{\mathrm M}_{k-1}$ depends on the TPM of the previous time steps through Eq.~\eqref{eq:cbarM}, every statistic computed inside the IMM depends on the past TPM, and Eq.~\eqref{eq:adapt} forms a feedback loop through the interaction step (dashed path in Fig.~\ref{fig:arch}).
The loop can lead to false declarations, when the maneuver probability is driven above the coast level by the loop rather than by the measurements, and to missed detections, when the TPM remains close to a fixed TPM or when a loss of sensitivity raises the smallest detectable maneuver.
When $s_k$ is the mode probability, the effect of the loop on the maneuver probability depends on how much the update mapping $\mathcal T$ in Eq.~\eqref{eq:adapt} changes $\pon$ and $\poff$ for a given change of the mode probability, that is, on the gain of $\mathcal T$.
If the gain is large, the change of $\bar c^{\mathrm M}_{k}$ in Eq.~\eqref{eq:cbarM} outweighs the likelihood ratio of the two models in Eq.~\eqref{eq:imm}, and $\mu^{\mathrm M}_k$ exceeds 0.5 without a maneuver.
If the gain is small, the TPM remains close to the initial value after a small maneuver, for which the mode probability changes little.
The predicted maneuver probability in Eq.~\eqref{eq:cbarM} then stays close to that of a fixed TPM, and the trade-off between coast accuracy and responsiveness to a maneuver remains.
In contrast, when $s_k$ is the NIS of the coast sub-filter or the likelihood computed from it, the mixing step of Eq.~\eqref{eq:mixC} pulls the coast sub-filter toward the maneuver sub-filter after a maneuver.
Consequently, the innovation of the coast sub-filter remains small, and the increase of $\pon$ after the maneuver is suppressed.
When the likelihood is computed before mixing~\cite{mondal2026maneuver}, the coast sub-filter is pulled in the same way, because it was updated from the mixed initial estimate at the previous time step.
The trade-off and the effect of the mixing on the coast sub-filter are examined in simulation in Section~\ref{sec:res_mech}.

\subsection{TPM adaptation driven by a decoupled reference filter}\label{sec:decoupled}\label{sec:ref}
In the proposed approach, $s_k$ is computed from a reference filter, as shown in Fig.~\ref{fig:arch}.
Following the threshold-based detection method of Goff et al.~\cite{goff2015tracking}, in which a coast filter declares a maneuver when its NIS exceeds a threshold, the reference filter is defined as a CKF with the same dynamics model, Eq.~\eqref{eq:coast}, and process noise $\tilde Q_{\mathrm C}$ as the coast sub-filter of the IMM.
The reference filter receives the same measurements as the IMM but is decoupled from the IMM.
The reference filter is updated with $\zv_k$ before the interaction step at $t_k$, and $\Pi_k$ is computed from $s_k$.
In Eq.~\eqref{eq:adapt}, the statistic $s_{k-1}$ computed inside the IMM at $t_{k-1}$ is available only after the IMM update at $t_{k-1}$ and therefore drives $\Pi_k$, whereas $s_k$ of the reference filter is available before the interaction step at $t_k$, since the reference filter does not use the output of the IMM at $t_k$.
Since $s_k$ is computed from the innovation of the reference filter, whose recursion uses neither $\Pi_k$ nor $\mu^j_k$, $s_k$ does not depend on the TPM or the mode probabilities of the IMM.
The feedback loop of Section~\ref{sec:loop} is therefore not formed.
The IMM affects the reference filter only through the covariance reset of Eq.~\eqref{eq:reset} immediately after the IMM declares a maneuver by Eq.~\eqref{eq:declare}.

The NIS of the reference filter at $t_k$ with $m_k$ available sensors is
\begin{equation}
    \Psiref_k=\nuv^{\mathrm{ref}\top}_k\big(S^{\mathrm{ref}}_k\big)^{-1}\nuv^{\mathrm{ref}}_k,\qquad
    \Psiref_k\mathrel{\dot\sim}\chi^2_{2m_k}\ \text{ during coast},
    \label{eq:psiref}
\end{equation}
where $\nuv^{\mathrm{ref}}_k$ and $S^{\mathrm{ref}}_k$ are the innovation and the innovation covariance of the reference filter computed by Eq.~\eqref{eq:ckfupd}.
During coast, $\nuv^{\mathrm{ref}}_k$ is approximately zero-mean Gaussian with covariance $S^{\mathrm{ref}}_k$, and $\Psiref_k$ therefore approximately follows the $\chi^2$ distribution with $2m_k$ degrees of freedom~\cite{bar2001estimation} when $\tilde Q_{\mathrm C}$ is calibrated to the unmodeled accelerations.
The effect of mis-calibration is examined in Section~\ref{sec:res_det}.
Since the reference filter follows the coast dynamics, it diverges from the target after a maneuver, and $\Psiref_k$ keeps increasing.
To restore the reference filter, the covariance is reset whenever $\Psiref_k$ exceeds the threshold $\Psi_{\max}$ of~\cite{goff2015tracking} or the IMM declares a maneuver:
\begin{equation}
    P^{\mathrm{ref}}_{k|k}=
    \begin{cases}
        P_0, & \Psiref_k>\Psi_{\max}\ \ \text{or}\ \ \mu^{\mathrm M}_{k}>0.5\ge\mu^{\mathrm M}_{k-1},\\
        P^{\mathrm{ref}}_{k|k-1}-K^{\mathrm{ref}}_kS^{\mathrm{ref}}_kK^{\mathrm{ref}\top}_k, & \text{otherwise},
    \end{cases}
    \label{eq:reset}
\end{equation}
where $P_0$ is the initial covariance, and the second reset condition corresponds to the IMM declaring a maneuver at $t_{k}$.
The state estimate $\hat\xv^{\mathrm{ref}}_{k|k}$ is not changed by the reset.
With the reset covariance, $\Psiref_k$ returns to the coast level within a few time steps, and the reference filter re-converges to the post-maneuver orbit.
The threshold $\Psi_{\max}$ is set equal to the declaration threshold of~\cite{goff2015tracking} but is used in \rdimm{} only for the reset, and $\Psiref_k$ itself is mapped continuously to the TPM as described in Section~\ref{sec:map}.
The second reset condition in Eq.~\eqref{eq:reset} addresses small maneuvers for which $\Psiref_k$ stays below $\Psi_{\max}$ and the first condition does not reset the covariance.
The IMM can declare a maneuver for which $\Psiref_k$ stays below $\Psi_{\max}$, because $\pon$ of Eq.~\eqref{eq:sigmoid} increases with $s_k$ regardless of $\Psi_{\max}$.
Without a reset, the reference filter keeps the small covariance of the coast phase, and the resulting small Kalman gain corrects the deviation of the state estimate caused by the maneuver only slightly.
The position error of the reference filter then increases continuously, and $\Psiref_k$ rises again after the declaration, which raises $\pon$ through the mapping of Section~\ref{sec:map} and can cause a second declaration of the same maneuver.
To prevent the second declaration, the second reset condition resets the covariance immediately after each declaration of the IMM.
Since the second reset condition applies only after a declaration, $\Psiref_k$ before the declaration is not affected.

\subsection{Continuous mapping to the transition probability}\label{sec:map}
The coast-to-maneuver probability $\ponk$ is computed from the NIS $\Psiref_k$ of the reference filter in three steps: a $\chi^2$ surprisal, a fading maximum and a logistic function.
The mapping is designed to keep $\pon$ small during coast and to increase $\pon$ gradually with the evidence of a maneuver in $\Psiref_k$ instead of at a hard threshold, while the false-alarm probability of the reference filter is kept as a design value.
As the number of available sensors $m_k$ and hence the number of degrees of freedom of $\Psiref_k$ in Eq.~\eqref{eq:psiref} change over time, $\Psiref_k$ is converted to a quantity that follows a single distribution during coast for every $m_k$.
Let $F_n$ denote the cumulative distribution function of the $\chi^2$ distribution with $n$ degrees of freedom.
The surprisal $s_k$, the negative logarithm of a tail probability~\cite{shannon1948mathematical,greenland2019valid}, is defined as
\begin{equation}
    s_k=\min\Big\{s_{\max},\ -\log_{10}\big[1-F_{2m_k}\big(\Psiref_k\big)\big]\Big\},
    \label{eq:surprise}
\end{equation}
where $1-F_{2m_k}(\Psiref_k)$ is the probability that a coast NIS exceeds the observed $\Psiref_k$.
If $\Psiref_k$ follows the $\chi^2$ distribution with $2m_k$ degrees of freedom during coast, the tail probability is uniformly distributed on $(0,1)$, and $P(s_k>c)=10^{-c}$ for every threshold $c<s_{\max}$ and every $m_k$.
Therefore, the surprisal inherently accounts for the change of the number of available sensors without a separate threshold for each $m_k$.
The upper bound $s_{\max}$ keeps $s_k$ finite for a very large NIS, and $s_k$ returns to the coast level within a few time steps after the reset of Eq.~\eqref{eq:reset}.

The maneuver indicator $\alpha_k$ is obtained from $s_k$ by a fading maximum, which retains a large value of $s_k$ over several time steps,
\begin{equation}
    \alpha_k=\max\big(\rho\,\alpha_{k-1},\ s_k/c\big),\qquad \alpha_0=0,
    \label{eq:alpha}
\end{equation}
where $\rho\in(0,1)$ is the fading factor and $c$ is the mapping constant.
After a large $s_k$, $\alpha_k$ decreases by the factor $\rho$ at each time step, and $s_{\max}$ limits the time for which $\alpha_k$ stays above $1$.
The fading maximum keeps $\pon$ raised for several measurement times after a large innovation.

\begin{figure}[!tb]\centering
    \includegraphics[width=\colfigwidth]{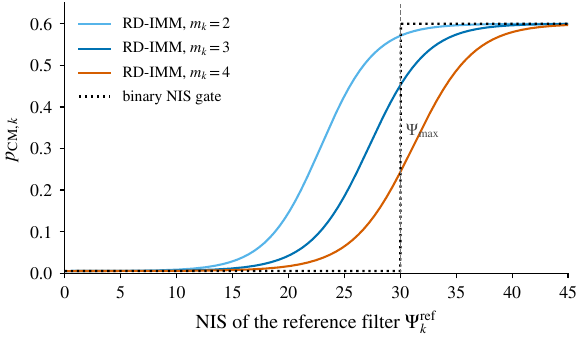}
    \caption{Mapping from the NIS of the reference filter to the coast-to-maneuver probability.}
    \label{fig:map}
\end{figure}

The coast-to-maneuver probability is then given by a logistic function of $\alpha_k$ centered at $\alpha_k=1$,
\begin{equation}
    \ponk=p_{\mathrm{lo}}+\frac{p_{\mathrm{hi}}-p_{\mathrm{lo}}}{1+\exp\big[{-\gamma(\alpha_k-1)}\big]},
    \label{eq:sigmoid}
\end{equation}
where $p_{\mathrm{lo}}$ and $p_{\mathrm{hi}}$ are the lower and upper bounds of $\pon$, and $\gamma$ is the steepness of the logistic function.
The logistic function keeps $\pon$ close to $p_{\mathrm{lo}}$ while the recent innovations are small and close to $p_{\mathrm{hi}}$ after a large innovation.
As $\gamma\to\infty$, the logistic function approaches a step function at $\alpha_k=1$, and $\pon$ switches between $p_{\mathrm{lo}}$ and $p_{\mathrm{hi}}$ at $s_k=c$, a threshold on the tail probability $10^{-c}$.
Figure~\ref{fig:map} shows $\ponk$ as a function of $\Psiref_k$ for $\alpha_{k-1}=0$ with two, three and four sensors.
For $c=\nCstar$, $s_k=c$ corresponds to $\Psiref_k=22.9$, 27.2 and 31.1 with two, three and four sensors, whereas the binary NIS gate of Section~\ref{sec:baselines} switches at $\Psi_{\max}=30$ for every $m_k$.
The mapping through the surprisal keeps the false-alarm probability independent of the number of available sensors and increases $\pon$ before $\Psiref_k$ reaches the threshold, which a hard threshold on $\Psiref_k$ does not provide.

As the logistic function of Eq.~\eqref{eq:sigmoid} is centered at $\alpha_k=1$, $\alpha_k$ reaching $1$ is regarded as a maneuver detection of the reference filter.
Since $\alpha_0=0$ and $\rho<1$, $\alpha_k$ first reaches $1$ at a time step with $s_k\ge c$.
During coast, $s_k\ge c$ occurs at each time step with a probability of approximately $10^{-c}$.
Over a coast interval with $N$ measurement times $t_1,\dots,t_N$, the union bound gives
\begin{equation}
    \Pr\Big\{\max_{k\le N}s_k\ge c\Big\}\lesssim N\,10^{-c}.
    \label{eq:union}
\end{equation}
The inequality holds without assuming that the $s_k$ are independent and remains valid under the reset of Eq.~\eqref{eq:reset} during coast, as $P_0$ is larger than the error covariance of the converged reference filter and the reset therefore increases $S^{\mathrm{ref}}_k$.
Setting the right-hand side of Eq.~\eqref{eq:union} to the allowed false-alarm probability $P_{\mathrm{FA}}$ of the reference filter gives
\begin{equation}
    c=\log_{10}\big(N/P_{\mathrm{FA}}\big),
    \label{eq:crule}
\end{equation}
where $N$ is the number of measurement times in the interval for which $P_{\mathrm{FA}}$ is specified.
For a given $P_{\mathrm{FA}}$, a longer interval therefore gives a larger $c$ and a smaller probability $10^{-c}=P_{\mathrm{FA}}/N$ at each time step.

The maneuver-to-coast probability is kept constant at $\poffk=\poff$.
The return of the IMM to the coast mode after a maneuver is then determined by the model likelihoods.
For a fixed $\pon$, the stationary distribution of the two-state Markov chain with the TPM of Eq.~\eqref{eq:tpmdef} has the maneuver probability $\pon/(\pon+\poff)$.
The stationary maneuver probability is $p_{\mathrm{lo}}/(p_{\mathrm{lo}}+\poff)$ when $\ponk=p_{\mathrm{lo}}$ and $p_{\mathrm{hi}}/(p_{\mathrm{hi}}+\poff)$ when $\ponk=p_{\mathrm{hi}}$.
The adapted TPM thus acts as a fixed TPM with $\pon=p_{\mathrm{lo}}$ during coast and with $\pon=p_{\mathrm{hi}}$ after a large innovation, which mitigates the trade-off in the choice of a fixed $\pon$ described in Section~\ref{sec:intro}.
The adapted TPM alone does not determine the declaration of the IMM by Eq.~\eqref{eq:declare}, because $\mu^{\mathrm M}_k$ in Eq.~\eqref{eq:imm} also depends on the model likelihoods.
The false-declaration rate of the IMM therefore differs from $P_{\mathrm{FA}}$ of the reference filter and is evaluated in Section~\ref{sec:res_det}.

\section{Simulation results and analysis}\label{sec:sim}

\newcommand{\figFailure}{%
\begin{widefigure}[!t]\centering
    \includegraphics[width=\textwidth]{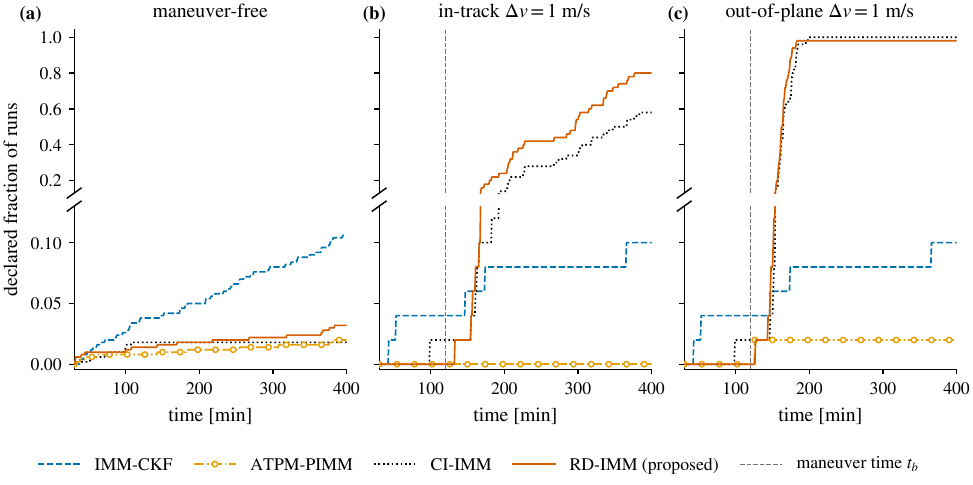}
    \caption{Cumulative fraction of runs that have declared a maneuver (\nMC{} Monte Carlo runs, 500 in the maneuver-free case).}
    \label{fig:declared}
\end{widefigure}}

\newcommand{\figMech}{%
\begin{widefigure}\centering
    \includegraphics[width=\textwidth]{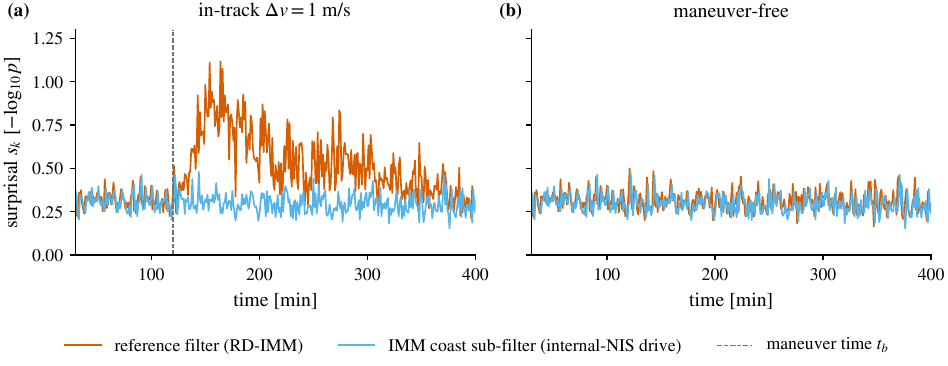}
    \caption{Surprisal of the reference filter and of the coast sub-filter of the internal-NIS drive (median of \nMC{} Monte Carlo runs).}
    \label{fig:mech}
\end{widefigure}}

\newcommand{\figRmse}{%
\begin{figure}[!tb]\centering
    \includegraphics[width=\colfigwidth]{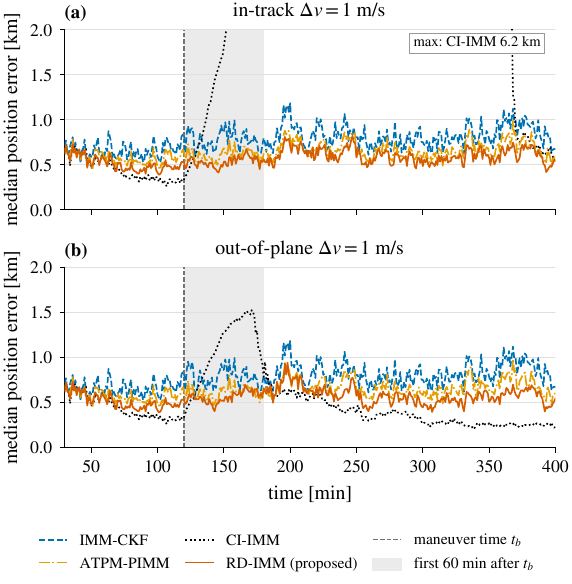}
    \caption{Position error for the \SI{1}{m/s} impulses (median of \nMC{} Monte Carlo runs).}
    \label{fig:rmse}
\end{figure}}

\newcommand{\tabResults}{%
\begin{widetable}\centering\footnotesize
\caption{Detection and tracking results (\nMC{} Monte Carlo runs per case, 500 maneuver-free runs for the false-declaration rate).}
\label{tab:results}
\setlength{\tabcolsep}{3.5pt}
\resizebox{\textwidth}{!}{%
\begin{tabular}{l cc cccc cccc c}
\toprule
& \multicolumn{2}{c}{maneuver-free} & \multicolumn{4}{c}{in-track $P_d$ / latency [min]} & \multicolumn{4}{c}{out-of-plane $P_d$ / latency [min]} & post-maneuver RMSE [km]\\
\cmidrule(lr){2-3}\cmidrule(lr){4-7}\cmidrule(lr){8-11}
Filter & false decl. & coast RMSE [km] & \SI{0.5}{m/s} & \SI{1}{m/s} & \SI{3}{m/s} & \SI{10}{m/s} & \SI{0.5}{m/s} & \SI{1}{m/s} & \SI{3}{m/s} & \SI{10}{m/s} & in-track \SI{1}{m/s}\\
\midrule
IMM-CKF & 0.11 & 1.09 & 0.06 / 54 & 0.06 / 54 & 0.16 / 8 & 1.00 / 4 & 0.06 / 54 & 0.06 / 54 & 0.12 / 17 & 0.98 / 4 & 1.10\\
ATPM-PIMM & 0.02 & 0.90 & 0.00 / -- & 0.00 / -- & 0.40 / 8 & 1.00 / 4 & 0.00 / -- & 0.02 / 6 & 0.50 / 9 & 1.00 / 4 & 0.88\\
CI-IMM & 0.02 & 0.45 & 0.02 / 199 & 0.56 / 151 & 0.98 / 15 & 1.00 / 5 & 0.14 / 108 & 0.98 / 45 & 0.98 / 13 & 1.00 / 5 & 2.25\\
\rdimm{} & 0.03 & 0.74 & 0.12 / 210 & 0.80 / 107 & 1.00 / 9 & 1.00 / 4 & 0.36 / 103 & 0.98 / 42 & 0.98 / 9 & 1.00 / 4 & 0.75\\
\midrule
\multicolumn{12}{l}{\emph{Variants of \rdimm}}\\
Internal-NIS drive & 0.02 & 0.74 & 0.00 / -- & 0.02 / 13 & 0.86 / 9 & 1.00 / 4 & 0.00 / -- & 0.02 / 7 & 0.92 / 9 & 1.00 / 4 & 0.73\\
Binary NIS gate & 0.04 & 0.72 & 0.02 / 199 & 0.56 / 128 & 0.98 / 9 & 1.00 / 4 & 0.14 / 108 & 0.98 / 45 & 0.98 / 9 & 1.00 / 4 & 0.72\\
\bottomrule
\end{tabular}}
\par\smallskip\parbox{\textwidth}{\raggedright\footnotesize --: no detected run.}
\end{widetable}
}

\subsection{Simulation setup}\label{sec:setup}
\begin{figure}[!b]\centering
    \includegraphics[width=\colfigwidth]{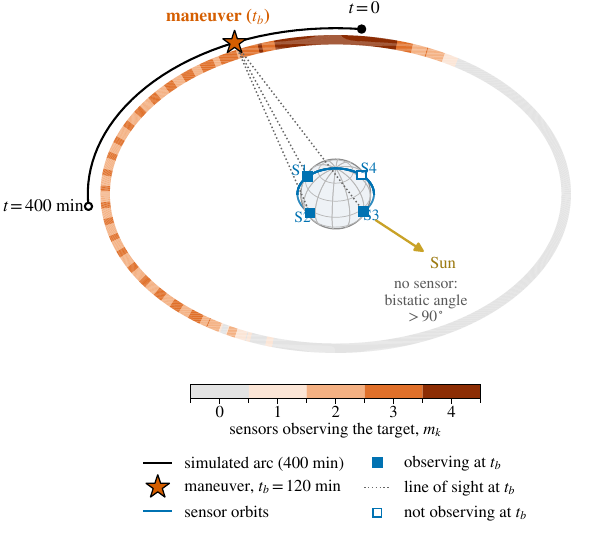}
    \caption{Space-based optical tracking scenario of a geostationary target.}
    \label{fig:scenario}
\end{figure}

The simulation scenario is adapted from Jia et al.~\cite{jia2015space}.
The target is observed by four space-based optical sensors with the angle-only measurement model of Section~\ref{sec:measmodel} (Fig.~\ref{fig:scenario}).
The four sensors are placed on a common circular orbit with an altitude of \SI{622}{km}, an inclination of \SI{96.0}{\degree} and a right ascension of the ascending node of \SI{306.0}{\degree}, and are spaced by \SI{90}{\degree} in the argument of latitude, starting from \SI{53}{\degree} at the initial epoch.
Each sensor measures the azimuth and elevation of the line of sight to the target in the ECI frame by Eq.~\eqref{eq:azel}.
A sensor provides a measurement when its line of sight is not blocked by the Earth and the bistatic solar angle at the target is below \SI{90}{\degree}, and its noise increases with the bistatic solar angle through $\kappa_q$.
The measurement noise of Eq.~\eqref{eq:azel} has the standard deviation $\sigma_\theta=\SI{0.0011}{\degree}$ per angle.
The measurements are taken every \SI{60}{s}, and two to four sensors observe the target at each measurement time.
The measurements of the available sensors are stacked as in Eq.~\eqref{eq:meas} instead of being fused by the covariance intersection of~\cite{jia2015space}, since the measurement noises of different sensors are independent.

The target is placed on the geostationary orbit at a longitude of approximately \SI{109.5}{\degree}E at the initial epoch.
The simulation starts at J2000 and lasts \SI{400}{min}, and the first \SI{30}{min} are excluded from every metric as filter warm-up.
The truth trajectory, Eq.~\eqref{eq:truth}, includes the zonal harmonics $J_2$--$J_6$, the lunisolar third-body accelerations and the solar radiation pressure with $C_RA/m=\SI{0.03}{m^2/kg}$~\cite{woo2021geostationary}.
The power spectral density of the process noise of the coast model is set to $\tilde Q_{\mathrm C}=\mathrm{diag}(\bm0_3,\,2.04\times10^{-14}\bm1_3)$~km$^2$/s$^3$ to account for the accelerations omitted in Eq.~\eqref{eq:PhiJ2}.
The discrete process noise of the maneuver model is $Q_{\mathrm M}=\mathrm{diag}(3\times10^{-6}\,\mathrm{km}^2\,\bm1_3,\,10^{-7}\,\mathrm{km}^2/\mathrm{s}^2\,\bm1_3)$ per integration step, which is 0.1 times the value of~\cite{jia2015space}.
The initial estimation error, which is common to all filters in a run, is drawn from a zero-mean Gaussian distribution with the covariance $P_0=\mathrm{diag}(100\,\mathrm{km}^2\,\bm1_3,\,10^{-2}\,\mathrm{km}^2/\mathrm{s}^2\,\bm1_3)$ of~\cite{jia2015space}, which is also used as the initial covariance of the filters.

A single impulse of magnitude $\dvs\in\{0.5,1,3,10\}$~m/s is applied at $t_b=\SI{120}{min}$,
either along the velocity (in-track) or along the ECI $z$ axis (out-of-plane), which correspond in direction to the east--west and north--south station keeping of a geostationary satellite~\cite{woo2021geostationary}.
Impulses of 0.1 and \SI{0.3}{m/s} are also applied to examine the detection floor.
For each impulse magnitude and direction, \nMC{} Monte Carlo runs with random measurement noise and initial error are performed, and the maneuver-free case is extended to 500 runs for the false-declaration rate.
In each run, all filters process the same measurements.
The simulation code was written in Python with the assistance of the AI coding agent Claude Code (Anthropic); the authors specified the filters and experiments and reviewed and verified the code and its outputs.

\earlyfloat{%
\begin{widetable}\centering\footnotesize
\caption{Detection and tracking results (\nMC{} Monte Carlo runs per case, 500 maneuver-free runs for the false-declaration rate).}
\label{tab:results}
\setlength{\tabcolsep}{3.5pt}
\resizebox{\textwidth}{!}{%
\begin{tabular}{l cc cccc cccc c}
\toprule
& \multicolumn{2}{c}{maneuver-free} & \multicolumn{4}{c}{in-track $P_d$ / latency [min]} & \multicolumn{4}{c}{out-of-plane $P_d$ / latency [min]} & post-maneuver RMSE [km]\\
\cmidrule(lr){2-3}\cmidrule(lr){4-7}\cmidrule(lr){8-11}
Filter & false decl. & coast RMSE [km] & \SI{0.5}{m/s} & \SI{1}{m/s} & \SI{3}{m/s} & \SI{10}{m/s} & \SI{0.5}{m/s} & \SI{1}{m/s} & \SI{3}{m/s} & \SI{10}{m/s} & in-track \SI{1}{m/s}\\
\midrule
IMM-CKF & 0.11 & 1.09 & 0.06 / 54 & 0.06 / 54 & 0.16 / 8 & 1.00 / 4 & 0.06 / 54 & 0.06 / 54 & 0.12 / 17 & 0.98 / 4 & 1.10\\
ATPM-PIMM & 0.02 & 0.90 & 0.00 / -- & 0.00 / -- & 0.40 / 8 & 1.00 / 4 & 0.00 / -- & 0.02 / 6 & 0.50 / 9 & 1.00 / 4 & 0.88\\
CI-IMM & 0.02 & 0.45 & 0.02 / 199 & 0.56 / 151 & 0.98 / 15 & 1.00 / 5 & 0.14 / 108 & 0.98 / 45 & 0.98 / 13 & 1.00 / 5 & 2.25\\
\rdimm{} & 0.03 & 0.74 & 0.12 / 210 & 0.80 / 107 & 1.00 / 9 & 1.00 / 4 & 0.36 / 103 & 0.98 / 42 & 0.98 / 9 & 1.00 / 4 & 0.75\\
\midrule
\multicolumn{12}{l}{\emph{Variants of \rdimm}}\\
Internal-NIS drive & 0.02 & 0.74 & 0.00 / -- & 0.02 / 13 & 0.86 / 9 & 1.00 / 4 & 0.00 / -- & 0.02 / 7 & 0.92 / 9 & 1.00 / 4 & 0.73\\
Binary NIS gate & 0.04 & 0.72 & 0.02 / 199 & 0.56 / 128 & 0.98 / 9 & 1.00 / 4 & 0.14 / 108 & 0.98 / 45 & 0.98 / 9 & 1.00 / 4 & 0.72\\
\bottomrule
\end{tabular}}
\par\smallskip\parbox{\textwidth}{\raggedright\footnotesize --: no detected run.}
\end{widetable}
}
\earlyfloat{\figFailure}

\subsection{Compared filters and metrics}\label{sec:baselines}
Three baseline filters are compared with \rdimm{}: the IMM-CKF of~\cite{jia2015space}, which is a standard IMM with a fixed TPM, ATPM-PIMM~\cite{xie2019adaptive} with an adaptive TPM, and the CI-IMM~\cite{goff2015tracking}, which switches from a single coast filter to an IMM of covariance-inflated models when the NIS of the coast filter exceeds a threshold~\cite{ru2009detection}.

The IMM-CKF consists of two 6-state CKFs of Eq.~\eqref{eq:coast}, one with $\tilde Q_{\mathrm C}$ and one with the discrete process noise $\mathrm{diag}(3\times10^{-5}\,\mathrm{km}^2\,\bm1_3,\,10^{-6}\,\mathrm{km}^2/\mathrm{s}^2\,\bm1_3)$ of~\cite{jia2015space} per step.
Its TPM, $\pon=0.03$ and $\poff=0.15$, is chosen manually since~\cite{jia2015space} does not report it.

ATPM-PIMM runs a fixed-TPM IMM in parallel with an adaptive IMM.
ATPM-PIMM is implemented with the coast and maneuver models of \rdimm{}, and the TPM of the fixed-TPM IMM and the initial TPM of the adaptive IMM are also identical to the TPM of the IMM-CKF.
In ATPM-PIMM, the TPM of the adaptive IMM is corrected with the mode-probability increments of the adaptive IMM.
When the likelihood ratio of the matching mode, which is used as the model-jump statistic, falls below the model jumping threshold $T_h$, the increments of the fixed-TPM IMM are used instead.
The threshold is set to $T_h=0.8$, within the range 0.7--0.9 recommended in~\cite{xie2019adaptive}.

The CI-IMM is implemented by replacing the EKF of~\cite{goff2015tracking} with the reference filter of Section~\ref{sec:ref}, and the other parameters of~\cite{goff2015tracking} are retained.
A maneuver is declared when $\Psiref_k>\Psi_{\max}=30$, and the estimate is then copied to 23 models whose covariances are multiplied by 10 until their traces reach limits from $10^{-3}$ to $10^{8}\,\mathrm{km}^2$.
The 23 models, which have the coast dynamics, are combined with the identity TPM, and a model whose mode probability falls below $10^{-10}$ is discarded until one model remains.
During the IMM phase, the NIS is the mean of the NIS of the models weighted by their mode probabilities, as in~\cite{goff2015tracking}.

\rdimm{} uses $p_{\mathrm{lo}}=0.005$, $p_{\mathrm{hi}}=0.60$, $\gamma=8$, $\poff=0.10$, $\rho=0.7$ and $s_{\max}=30$, and $c=\nCstar$ is obtained from Eq.~\eqref{eq:crule} with $P_{\mathrm{FA}}=0.05$ for the $N=\nNtestGeo$ measurement times after the warm-up.
To examine the effect of the source of the NIS and of the continuous mapping, two variants of \rdimm{} are also evaluated.
The first variant, the internal-NIS drive, replaces $\Psiref_k$ in Eq.~\eqref{eq:surprise} with the NIS $\Psi^{\mathrm C}_k$ of the coast sub-filter of the IMM in Eq.~\eqref{eq:nis}, forming the loop of Section~\ref{sec:loop}.
The second variant, the binary NIS gate, replaces $s_k/c$ in Eq.~\eqref{eq:alpha} with $\alpha_{\mathrm{hi}}=3$ when $\Psiref_k>\Psi_{\max}$ and with $0$ otherwise (Fig.~\ref{fig:map}).
The binary NIS gate represents detection-driven TPM adaptation, which Mondal et al.~\cite{mondal2026maneuver} consider to be delayed because the detection relies on a moving average of the NIS.
Since the gate uses the NIS at each measurement time without a moving average, the gate differs from \rdimm{} only in the binary threshold.
The threshold $\Psi_{\max}$ of the gate is the declaration threshold of the CI-IMM.

The filters are evaluated with the following metrics.
The CI-IMM is regarded as declaring a maneuver when its NIS exceeds $\Psi_{\max}$, and the other filters when Eq.~\eqref{eq:declare} is satisfied.
The detection probability $P_d$ is the fraction of runs in which a maneuver is declared after $t_b$, including declarations not caused by the maneuver, and $P_d$ is therefore compared with the false-declaration rate.
The false-declaration rate is the fraction of maneuver-free runs in which a maneuver is declared at any time after warm-up.
The latency is the median time from $t_b$ to the first declaration among the detected runs.
The coast RMSE is the position RMSE over the maneuver-free runs after warm-up, and the post-maneuver RMSE is the position RMSE over the first \SI{60}{min} after $t_b$.
The false-declaration rate is computed from the 500 maneuver-free runs, and the other metrics and the parameter variations from the first \nMC{} runs.

\subsection{Effect of the decoupled reference filter and the continuous mapping}\label{sec:res_mech}\label{sec:res_fail}

The TPM adaptation of \rdimm{} mitigates the trade-off of a fixed TPM between coast accuracy and responsiveness to a maneuver (Section~\ref{sec:intro}).
\rdimm{} has a coast RMSE of \nPropRmseCoast{}~km, compared with \nJiaRmseCoast{} and \SI{0.90}{km} for the IMM-CKF and ATPM-PIMM (Table~\ref{tab:results}).
\latefloat{%
\begin{widetable}\centering\footnotesize
\caption{Detection and tracking results (\nMC{} Monte Carlo runs per case, 500 maneuver-free runs for the false-declaration rate).}
\label{tab:results}
\setlength{\tabcolsep}{3.5pt}
\resizebox{\textwidth}{!}{%
\begin{tabular}{l cc cccc cccc c}
\toprule
& \multicolumn{2}{c}{maneuver-free} & \multicolumn{4}{c}{in-track $P_d$ / latency [min]} & \multicolumn{4}{c}{out-of-plane $P_d$ / latency [min]} & post-maneuver RMSE [km]\\
\cmidrule(lr){2-3}\cmidrule(lr){4-7}\cmidrule(lr){8-11}
Filter & false decl. & coast RMSE [km] & \SI{0.5}{m/s} & \SI{1}{m/s} & \SI{3}{m/s} & \SI{10}{m/s} & \SI{0.5}{m/s} & \SI{1}{m/s} & \SI{3}{m/s} & \SI{10}{m/s} & in-track \SI{1}{m/s}\\
\midrule
IMM-CKF & 0.11 & 1.09 & 0.06 / 54 & 0.06 / 54 & 0.16 / 8 & 1.00 / 4 & 0.06 / 54 & 0.06 / 54 & 0.12 / 17 & 0.98 / 4 & 1.10\\
ATPM-PIMM & 0.02 & 0.90 & 0.00 / -- & 0.00 / -- & 0.40 / 8 & 1.00 / 4 & 0.00 / -- & 0.02 / 6 & 0.50 / 9 & 1.00 / 4 & 0.88\\
CI-IMM & 0.02 & 0.45 & 0.02 / 199 & 0.56 / 151 & 0.98 / 15 & 1.00 / 5 & 0.14 / 108 & 0.98 / 45 & 0.98 / 13 & 1.00 / 5 & 2.25\\
\rdimm{} & 0.03 & 0.74 & 0.12 / 210 & 0.80 / 107 & 1.00 / 9 & 1.00 / 4 & 0.36 / 103 & 0.98 / 42 & 0.98 / 9 & 1.00 / 4 & 0.75\\
\midrule
\multicolumn{12}{l}{\emph{Variants of \rdimm}}\\
Internal-NIS drive & 0.02 & 0.74 & 0.00 / -- & 0.02 / 13 & 0.86 / 9 & 1.00 / 4 & 0.00 / -- & 0.02 / 7 & 0.92 / 9 & 1.00 / 4 & 0.73\\
Binary NIS gate & 0.04 & 0.72 & 0.02 / 199 & 0.56 / 128 & 0.98 / 9 & 1.00 / 4 & 0.14 / 108 & 0.98 / 45 & 0.98 / 9 & 1.00 / 4 & 0.72\\
\bottomrule
\end{tabular}}
\par\smallskip\parbox{\textwidth}{\raggedright\footnotesize --: no detected run.}
\end{widetable}
\unskip}%
\latefloat{\figFailure}%
At the same time, \rdimm{} detects the in-track and out-of-plane \SI{1}{m/s} impulses with $P_d=\nPropPdTone$ and \nPropPdZone{}, compared with \nJiaPdTone{} and 0.06 for the IMM-CKF and \nXiePdTone{} and \nXiePdZone{} for ATPM-PIMM.
For the IMM-CKF with $\pon=0.03$, $P_d$ for the in-track and out-of-plane \SI{1}{m/s} impulses is below the false-declaration rate of \nJiaFa{} (Fig.~\ref{fig:declared}a--c).
When $\pon$ of the IMM-CKF is varied from 0.001 to 0.3, the coast RMSE increases from 0.71 to \SI{1.46}{km}, whereas $P_d$ for the in-track and out-of-plane \SI{1}{m/s} impulses does not exceed the false-declaration rate for any $\pon$.
For ATPM-PIMM, $P_d$ for the in-track and out-of-plane \SI{1}{m/s} impulses does not exceed the false-declaration rate of \nXieFa{}.
In the first \SI{60}{min} after the in-track and out-of-plane \SI{1}{m/s} impulses, the largest $\pon$ of the adaptive IMM of ATPM-PIMM in each run has a median of 0.037, close to the initial value of 0.03, as expected for a small gain of the TPM update (Section~\ref{sec:loop}).
Since ATPM-PIMM uses the same models as \rdimm{} (Section~\ref{sec:baselines}), the higher detection probability of \rdimm{} originates from the TPM adaptation driven by the reference filter.
The higher detection probability of \rdimm{} is also obtained with a 9-state constant-acceleration maneuver model.
The 9-state model adds the acceleration to the state of the maneuver model, with the power spectral density $\mathrm{diag}(\bm0_3,\,2.04\times10^{-14}\bm1_3,\,3.75\times10^{-12}\bm1_3)$ in km$^2$/s$^3$ and km$^2$/s$^5$ and an initial acceleration of zero with a standard deviation of \SI{4.5e-5}{km/s^2} per axis.
With the 9-state model, \rdimm{} detects 0.80 of the in-track and 0.98 of the out-of-plane \SI{1}{m/s} impulses, as with the 6-state model, with a false-declaration rate of 0.06 and a coast RMSE of \SI{0.96}{km} in the first \nMC{} maneuver-free runs.
A fixed TPM with the 9-state model detects 0.20 of the impulses, close to its false-declaration rate of 0.24.

The reference filter keeps the sensitivity to small maneuvers, because the reference filter is decoupled from the IMM.
The effect of the decoupling is examined with the internal-NIS drive, which differs from \rdimm{} only in that the NIS is taken from the coast sub-filter of the IMM.
In the maneuver-free runs, the surprisals of the two filters have medians of $0.31$ and $0.29$, close to $-\log_{10}0.5\approx0.30$ (Fig.~\ref{fig:mech}b), which is the median of $s_k$ for a tail probability uniformly distributed on $(0,1)$ (Section~\ref{sec:map}).
After the in-track \SI{1}{m/s} impulse, the median surprisal of the reference filter reaches $1.12$ at \SI{44}{min} after the maneuver, whereas the median surprisal of the coast sub-filter remains near $0.3$ (Fig.~\ref{fig:mech}a).
In the internal-NIS drive, the mixing of Eq.~\eqref{eq:mixC} pulls the coast sub-filter toward the maneuver estimate and reduces the NIS of the coast sub-filter (Section~\ref{sec:loop}).
Consequently, the internal-NIS drive detects the in-track and out-of-plane \SI{1}{m/s} impulses with $P_d=\nOwnPdTone$ and 0.02 at a false-declaration rate of \nOwnFa{}, compared with \nPropPdTone{} and \nPropPdZone{} for \rdimm{} (Table~\ref{tab:results}).
\earlyfloat{\figMech}
\latefloat{\figMech}

The continuous mapping of $\Psiref_k$ to $\pon$ increases the detection probability of the maneuvers for which $\Psiref_k$ remains below $\Psi_{\max}$ (Section~\ref{sec:map}).
The binary NIS gate raises $\pon$ only when $\Psiref_k$ exceeds $\Psi_{\max}$, whereas \rdimm{} raises $\pon$ continuously with $s_k$.
At similar false-declaration rates of \nGateFa{} and \nPropFaNull{}, the gate detects the in-track \SI{1}{m/s} impulse with $P_d=\nGatePdTone$ and a median latency of \nGateLatTone{}~min.
\rdimm{} detects the same impulse with $P_d=\nPropPdTone$ and a median latency of \nPropLatTone{}~min.
Of the 40 runs in which \rdimm{} detects the in-track \SI{1}{m/s} impulse, 20 declare the maneuver before $\Psiref_k$ exceeds $\Psi_{\max}$.
In 13 of the 20 runs, $\Psiref_k$ remains below $\Psi_{\max}$ and the gate declares no maneuver.
For the out-of-plane \SI{1}{m/s} impulse, $\Psiref_k$ exceeds $\Psi_{\max}$ in 49 of the \nMC{} runs, and the gate detects the maneuver in the same runs as \rdimm{}, with median latencies of \nGateLatZone{} and \nPropLatZone{}~min.

\subsection{Detection performance}\label{sec:res_det}
\earlyfloat{\figRmse}
The IMM-CKF and ATPM-PIMM detect the maneuver in almost every run only at \SI{10}{m/s}.
Below \SI{10}{m/s}, \rdimm{} is therefore compared with the CI-IMM, whose threshold-based detection method motivates the reference filter of \rdimm{} (Section~\ref{sec:ref}).
\rdimm{} has a higher detection probability for the out-of-plane \SI{0.5}{m/s}, in-track \SI{0.5}{m/s} and in-track \SI{1}{m/s} maneuvers, with $P_d=\nPropPdZhalf$, \nPropPdThalf{} and \nPropPdTone{} compared with 0.14, 0.02 and \nGoffPdTone{} for the CI-IMM, and both filters detect the out-of-plane \SI{1}{m/s} and the \SI{3}{m/s} maneuvers in at least 49 of the \nMC{} runs (Fig.~\ref{fig:declared}b,c).
In the runs detected by both filters, which are 27 runs for the in-track \SI{1}{m/s} maneuver, 49 runs for the out-of-plane \SI{1}{m/s} maneuver and 49 runs for each \SI{3}{m/s} maneuver, the median latencies of \rdimm{} are 88, 42 and \SI{9}{min}, compared with 148, 45 and 13--\SI{15}{min} for the CI-IMM.
Both filters detect the \SI{10}{m/s} maneuvers in every run with median latencies of 4--\SI{5}{min}, and impulses of 0.1 and \SI{0.3}{m/s} are rarely detected by any of the compared filters.
The detection floor of \rdimm{} in this scenario is therefore approximately 0.5--\SI{1}{m/s}, below the velocity change of 1--2~m/s of a north--south station-keeping maneuver~\cite{woo2021geostationary}.

Over 500 maneuver-free runs, the false-declaration rate of \rdimm{} is $0.032$, below the level $P_{\mathrm{FA}}=0.05$ set for the reference filter by Eq.~\eqref{eq:crule}, compared with $0.018$ for the CI-IMM, $0.106$ for the IMM-CKF and $0.020$ for ATPM-PIMM.
The CI-IMM has a lower false-declaration rate than \rdimm{} because the threshold $\Psi_{\max}=30$ gives approximately 0.02 by the bound of Eq.~\eqref{eq:union} over the $N$ measurement times with two to four sensors, whereas $c$ of \rdimm{} is set by Eq.~\eqref{eq:crule} for $P_{\mathrm{FA}}=0.05$.
A smaller $c$ shortens the latency at the cost of more false declarations.
With $c=3$, the in-track \SI{1}{m/s} latency decreases from \nPropLatTone{} to \SI{52}{min} ($P_d=0.96$), and the false-declaration rate in the first \nMC{} maneuver-free runs increases from \nPropFaNullFifty{} to $0.22$.
With $c=5$, no maneuver-free run declares a maneuver, and the detection probability at \SI{1}{m/s} decreases to $0.18$ in-track and $0.42$ out-of-plane.

\rdimm{} also detects more of the in-track \SI{1}{m/s} maneuvers than the CI-IMM with a mis-calibrated $\tilde Q_{\mathrm C}$.
The power spectral density $\tilde Q_{\mathrm C}$ is scaled by 0.1 and 10 in the reference filter and in the coast models of \rdimm{} and the CI-IMM.
With both factors, the false-declaration rates of the two filters in the first \nMC{} runs remain \nPropFaNullFifty{}.
With the factors 0.1 and 10, \rdimm{} detects 0.94 and 0.28 of the in-track \SI{1}{m/s} maneuvers, compared with 0.76 and 0.12 for the CI-IMM.

\subsection{Tracking accuracy}\label{sec:res_rmse}
\latefloat{\figRmse}
Figure~\ref{fig:rmse} presents the median position error for the \SI{1}{m/s} impulses, and Table~\ref{tab:results} lists the RMSE values.
\rdimm{} has a lower coast RMSE than the IMM-CKF and ATPM-PIMM (Section~\ref{sec:res_mech}), because $\pon$ of \rdimm{} remains below 0.01 at 97\% of the time steps after warm-up in the maneuver-free runs and the maneuver model therefore has a smaller weight in the mixing of Eq.~\eqref{eq:mixC}.
The coast RMSE of the IMM-CKF also decreases to \SI{0.71}{km} at $\pon=0.001$ (Section~\ref{sec:res_mech}).
After the 1--3~m/s maneuvers, the post-maneuver RMSE of \rdimm{} is 0.75--\SI{0.92}{km}, compared with 0.88--\SI{1.13}{km} for the IMM-CKF and ATPM-PIMM, whose post-maneuver RMSE remains close to their coast RMSE.
In \rdimm{}, the maneuver mode dominates only for a short time after the declaration, and $\Pca_k$ falls below 0.5 within a few minutes as the filter converges to the post-maneuver orbit.

\rdimm{} also keeps the error close to the coast level during the detection delay, because the IMM with the maneuver model runs at every time step and corrects the deviation before the declaration.
In the CI-IMM, the converged coast filter corrects the deviation caused by the impulse only slightly until the maneuver is declared.
The median position error of the CI-IMM reaches \SI{6.0}{km} two hours after the in-track \SI{1}{m/s} impulse (Fig.~\ref{fig:rmse}a).
For the out-of-plane \SI{1}{m/s} and the \SI{3}{m/s} impulses, the median error of the CI-IMM at the declaration is 1.54--\SI{2.56}{km}, compared with 0.55--\SI{0.62}{km} for \rdimm{}.
The post-maneuver RMSE of the CI-IMM after the 1--3~m/s maneuvers is 1.17--\SI{2.25}{km}, compared with 0.75--\SI{0.92}{km} for \rdimm{}.
After the re-convergence and in the maneuver-free runs, the CI-IMM is more accurate than \rdimm{} because it runs only the coast model (coast RMSE of \SI{0.45}{km}, Fig.~\ref{fig:rmse} and Table~\ref{tab:results}).

\subsection{Ablations and computation time}\label{sec:breadth}
\earlyfloat{\begin{widetable}\centering\footnotesize
\caption{Incorporation of the reference NIS into the IMM (50 Monte Carlo runs).}
\label{tab:inject}\setlength{\tabcolsep}{3pt}\resizebox{\textwidth}{!}{\begin{tabular}{l cc cccc c}\toprule
& \multicolumn{2}{c}{maneuver-free} & \multicolumn{2}{c}{in-track $P_d$ / latency [min]} & \multicolumn{2}{c}{out-of-plane $P_d$ / latency [min]} & post-maneuver\\\cmidrule(lr){2-3}\cmidrule(lr){4-5}\cmidrule(lr){6-7}
NIS incorporated as & false decl. & coast RMSE [km] & \SI{1}{m/s} & \SI{3}{m/s} & \SI{0.5}{m/s} & \SI{1}{m/s} & RMSE [km], in-track \SI{1}{m/s}\\\midrule
TPM (\rdimm{}) & 0.02 & 0.74 & 0.80 / 107 & 1.00 / 9 & 0.36 / 103 & 0.98 / 42 & 0.75\\
none & 0.00 & 0.72 & 0.02 / 13 & 0.88 / 9 & 0.00 / -- & 0.02 / 7 & 0.71\\
\addlinespace
likelihood, ratio 30 & 0.02 & 0.45 & 0.72 / 168 & 0.90 / 15 & 0.42 / 103 & 0.90 / 44 & 2.05\\
likelihood + TPM, ratio 300 & 0.04 & 0.45 & 0.62 / 177 & 0.54 / 16 & 0.42 / 103 & 0.70 / 45 & 2.19\\
\addlinespace
state, $\beta=1.25$ & 0.02 & 0.71 & 0.70 / 36 & 1.00 / 8 & 0.50 / 74 & 0.98 / 29 & 0.81\\
state, $\beta=1.5$ & 0.00 & 0.71 & 0.46 / 36 & 1.00 / 8 & 0.28 / 78 & 0.94 / 33 & 0.79\\
state, $\beta=2$ & 0.00 & 0.71 & 0.22 / 46 & 1.00 / 9 & 0.06 / 86 & 0.82 / 38 & 0.75\\
\bottomrule\end{tabular}}
\end{widetable}
}
In Table~\ref{tab:inject}, different ways of incorporating the NIS $\Psiref_k$ into the IMM are examined.
\latefloat{\begin{widetable}\centering\footnotesize
\caption{Incorporation of the reference NIS into the IMM (50 Monte Carlo runs).}
\label{tab:inject}\setlength{\tabcolsep}{3pt}\resizebox{\textwidth}{!}{\begin{tabular}{l cc cccc c}\toprule
& \multicolumn{2}{c}{maneuver-free} & \multicolumn{2}{c}{in-track $P_d$ / latency [min]} & \multicolumn{2}{c}{out-of-plane $P_d$ / latency [min]} & post-maneuver\\\cmidrule(lr){2-3}\cmidrule(lr){4-5}\cmidrule(lr){6-7}
NIS incorporated as & false decl. & coast RMSE [km] & \SI{1}{m/s} & \SI{3}{m/s} & \SI{0.5}{m/s} & \SI{1}{m/s} & RMSE [km], in-track \SI{1}{m/s}\\\midrule
TPM (\rdimm{}) & 0.02 & 0.74 & 0.80 / 107 & 1.00 / 9 & 0.36 / 103 & 0.98 / 42 & 0.75\\
none & 0.00 & 0.72 & 0.02 / 13 & 0.88 / 9 & 0.00 / -- & 0.02 / 7 & 0.71\\
\addlinespace
likelihood, ratio 30 & 0.02 & 0.45 & 0.72 / 168 & 0.90 / 15 & 0.42 / 103 & 0.90 / 44 & 2.05\\
likelihood + TPM, ratio 300 & 0.04 & 0.45 & 0.62 / 177 & 0.54 / 16 & 0.42 / 103 & 0.70 / 45 & 2.19\\
\addlinespace
state, $\beta=1.25$ & 0.02 & 0.71 & 0.70 / 36 & 1.00 / 8 & 0.50 / 74 & 0.98 / 29 & 0.81\\
state, $\beta=1.5$ & 0.00 & 0.71 & 0.46 / 36 & 1.00 / 8 & 0.28 / 78 & 0.94 / 33 & 0.79\\
state, $\beta=2$ & 0.00 & 0.71 & 0.22 / 46 & 1.00 / 9 & 0.06 / 86 & 0.82 / 38 & 0.75\\
\bottomrule\end{tabular}}
\end{widetable}
\unskip}%
The row ``none'' corresponds to the IMM of \rdimm{} with a fixed TPM, without $\Psiref_k$.
With the same fixed $\pon$, $\Psiref_k$ is incorporated as a model likelihood in the row ``likelihood, ratio 30'' and as a state in the rows ``state''.
In the row ``likelihood + TPM'', $\Psiref_k$ is incorporated as a likelihood and also drives the TPM as in \rdimm{}.
For the likelihood incorporation, $\Psiref_k$ is converted to the density ratio of $r\chi^2_{2m_k}$ to $\chi^2_{2m_k}$, given by $r^{-m_k}\exp[\Psiref_k(1-1/r)/2]$, and the likelihood of the maneuver model is multiplied by the ratio.
With $r=30$, the false-declaration rate and the coast RMSE are 0.02 and \SI{0.45}{km}, compared with 0.02 and \SI{0.74}{km} for \rdimm{}.
However, the in-track \SI{1}{m/s} maneuver is detected in 0.72 of the runs after \SI{168}{min}, compared with 0.80 after \SI{107}{min} for \rdimm{}.
The post-maneuver RMSE for the in-track \SI{1}{m/s} maneuver is \SI{2.05}{km}, compared with \SI{0.75}{km} for \rdimm{}.
For the state incorporation, the orbital state of both models is augmented with a scalar $\eta_k$, the expected value of $\Psiref_k/(2m_k)$, which is 1 without a maneuver.
$\eta_k$ starts at 1 and is constant in the coast model and a random walk with variance 0.01 per \SI{60}{s} in the maneuver model.
The quantity $\Psiref_k/(2m_k)=\eta_k+v_k$ with $\mathrm{var}(v_k)=\beta/m_k$ is processed as an additional measurement.
With $\beta=1.25$, the in-track \SI{1}{m/s} maneuver is detected in 0.70 of the runs after \SI{36}{min}, compared with 0.80 after \SI{107}{min} for \rdimm{}.
However, $\beta$ is selected empirically, whereas $c$ is determined from $P_{\mathrm{FA}}$ by Eq.~\eqref{eq:crule}.
With $\beta=1.25$ and measurement intervals of 5 and \SI{15}{min}, the in-track \SI{1}{m/s} maneuver is detected in 0.26 and 0.12 of the runs, compared with 0.82 and 0.52 for \rdimm{} with $c$ from Eq.~\eqref{eq:crule}.
The false-declaration rates of both filters with the measurement intervals of 5 and \SI{15}{min} are at most 0.04.
With $\beta=1.5$, the false-declaration rate over 500 maneuver-free runs is 0.038, close to 0.032 for \rdimm{}.
The in-track \SI{1}{m/s} maneuver is then detected in 0.46 of the runs, compared with 0.80 for \rdimm{}.

In a Python implementation on an Apple M2, the reference filter increases the computation time per measurement time from 2.76 to \SI{4.35}{ms}.
When the prediction of the reference filter is computed in parallel with the IMM update, the computation time per measurement time is \SI{3.04}{ms}.
The increase of the computation time is small compared with the \SI{60}{s} measurement interval.

\section{Conclusion}\label{sec:conclusion}
This paper examined adaptive-TPM IMM filters that drive the TPM from statistics computed inside the IMM.
The statistics depend recursively on the TPM through the interaction step, and the dependence forms a feedback loop.
The analysis of the interaction step indicates that, when the TPM is driven by the mode probability with a large gain of the TPM update, the loop can raise the maneuver probability in the absence of a maneuver.
In contrast, with a small gain of the TPM update, the TPM stays close to its initial value after a small maneuver, as observed for ATPM-PIMM in the simulation.
When the TPM is driven by the NIS of the coast sub-filter or the likelihood computed from it, the loop can suppress the rise of the maneuver probability after a maneuver, as observed for the internal-NIS drive.
To eliminate the loop, \rdimm{} takes the NIS from a reference filter decoupled from the IMM and maps it continuously to the coast-to-maneuver transition probability, with a mapping constant determined from the allowed false-alarm probability of the reference filter.
In a space-based optical scenario with a geostationary target, the reference filter kept its sensitivity after small maneuvers, and \rdimm{} mitigated the trade-off of a fixed TPM.
With the continuous mapping, \rdimm{} also declared maneuvers for which the NIS of the reference filter remained below the threshold of the CI-IMM.
\rdimm{} detected 1--3~m/s maneuvers, which include north--south station-keeping maneuvers, with a false-declaration rate below the false-alarm probability set for the reference filter.
In the first hour after the 1--3~m/s maneuvers, \rdimm{} also had a lower position error than the compared filters and avoided the error growth of the CI-IMM during the detection delay.
The simulation results were obtained with measurements every minute from dedicated tracking, and sparse observations separated by long gaps and low-thrust maneuvers remain for future work.

\section*{CRediT authorship contribution statement}
\textbf{Euiseok Han:} Conceptualization, Data curation, Formal analysis, Investigation, Methodology, Project administration, Software, Validation, Visualization, Writing -- original draft, Writing -- review \& editing.
\textbf{Sangheon Choi:} Methodology, Supervision, Validation, Writing -- review \& editing.
\textbf{Seung-Hyun Kong:} Supervision, Writing -- review \& editing.

\section*{Funding}
This research did not receive any specific grant from funding agencies in the public, commercial, or not-for-profit sectors.

\section*{Declaration of generative AI and AI-assisted technologies in the manuscript preparation process}
During the preparation of this work the authors used Claude (Anthropic) in order to assist with drafting and editing the manuscript text.
After using this tool, the authors reviewed and edited the content as needed and take full responsibility for the content of the published article.

\bibliographystyle{elsarticle-num-acta}
\bibliography{references}

\end{document}